# Dirac parameters and topological phase diagram of Pb$_{1-x}$Sn$_x$Se from magneto-spectroscopy

G. Krizman,[1] B.A. Assaf,[2,3] T. Phuphachong,[1] G. Bauer,[4] G. Springholz,[4] L.A. de Vaulchier,[1] Y. Guldner[1]

[1] Laboratoire Pierre Aigrain, Département de Physique, Ecole Normale Supérieure, PSL Research University, Sorbonne Université, CNRS, 24 rue Lhomond, 75005 Paris, France

[2] Département de Physique, Ecole Normale Supérieure, PSL Research University, CNRS, 24 rue Lhomond, 75005 Paris, France

[3] Department of Physics, University of Notre Dame, Notre Dame, IN 46556, USA

[4] Institut für Halbleiter- und Festkörperphysik, Johannes Kepler Universität, Linz, Altenberger Strasse, 69, 4040 Linz, Austria

**Pb$_{1-x}$Sn$_x$Se hosts 3D massive Dirac fermions across the entire composition range for which the crystal structure is cubic. In this work, we present a comprehensive experimental mapping of the 3D band structure parameters of Pb$_{1-x}$Sn$_x$Se as a function of composition and temperature. We cover a parameter space spanning the band inversion that yields its topological crystalline insulator phase. A non-closure of the energy gap is evidenced in the vicinity of this phase transition. Using magnetooptical Landau level spectroscopy, we determine the energy gap, Dirac velocity, anisotropy factor and topological character of Pb$_{1-x}$Sn$_x$Se epilayers grown by molecular beam epitaxy on BaF$_2$ (111). Our results are evidence that Pb$_{1-x}$Sn$_x$Se is a model system to study topological phases and the nature of the phase transition.**

Pb$_{1-x}$Sn$_x$Se is a narrow-gap semiconductor that hosts 3D massive Dirac fermions. [1] [2] [3] It is known to undergo a topological phase transition in the bulk that leads to a topological crystalline insulator (TCI). [4] [5] [6] [7] [8] The transition can be induced by increasing the Sn to Pb ratio or by cooling down the system. [6] [9] In addition to its tunability, Pb$_{1-x}$Sn$_x$Se is valley degenerate. The 3D Dirac fermions in Pb$_{1-x}$Sn$_x$Se occur at four $L$-points in the Brillouin zone. This is illustrated in Fig. 1(c-e) where an ellipsoidal Fermi surface occurring at one of the $L$-points has its long axis along the [111] direction and three others have their long axis tilted along the [$\bar{1}$11]. The first one is referred to as the longitudinal valley and the second ones are the oblique valleys. The band inversion therefore yields four Dirac cones per 2D Brillouin zone stemming from four bulk band inversions and protected by mirror symmetry. [5] [10] [11] [12] Because of the Fermi surface anisotropy, three of the Dirac cones are identical and occur at the $\bar{M}$ points and the fourth occurs at the $\bar{\Gamma}$ point of the 2D Brillouin zone (see Fig. 1(c)).

One highly relevant and interesting aspect of the IV-VI Pb$_{1-x}$Sn$_x$Se materials is their tunable character versus chemical composition and temperature. [9] [13] [14] [15] Although investigated in recent angle-resolved photoemission experiments (ARPES), the changing band structure parameters of bulk states has remain unaddressed because in ARPES the bulk band structure is largely masked by the topological surface states. [6] [14] [16] The band parameters are of key importance for band structure calculations and phenomenological models that aim to study the electronic, optical and thermoelectric properties of Pb$_{1-x}$Sn$_x$Se. [2] [17] [18] [19] [20] The parameters include the energy gap, the anisotropy factor, but also the changing Dirac velocity and band edge (Dirac) mass versus temperature and composition. Moreover, the variation of the band inversion point as a function of temperature and composition and its nature are decisive for the topological properties of the system.

In the present work, we thus systematically investigate the changing band structure parameters of Pb$_{1-x}$Sn$_x$Se versus composition and temperature. We use magnetooptical infrared spectroscopy to study high mobility (111)-oriented epilayers of Pb$_{1-x}$Sn$_x$Se grown by molecular beam epitaxy (MBE) on BaF$_2$ substrates. We probe the Landau level (LL) transitions in a series of samples having different compositions ($0 < x < 0.3$) at various temperatures (4.2 K $< T <$ 200 K). From the results, we derive the variation of the bulk energy gap, Dirac velocity, band-edge mass and anisotropy and thereby construct the topological phase diagram versus chemical composition and temperature. Most notable, we consistently observe that the bulk gap does not close to zero across the topological phase transition, which is not only of fundamental importance but also relevant for topological devices aiming at tuning the topological properties by external knobs like strain or applied electric fields. [21] [22] [23]

Three dimensional $Pb_{1-x}Sn_xSe$ TCI crystallizes in the rocksalt structure depicted in Fig. 1(a). These alloys present in a first approximation electron-hole symmetric band extrema at the 4 $L$-points of their Brillouin zone. In the trivial regime, $L_6^+$ ($L_6^-$) forms the valence (conduction) band as shown in Fig. 1(b). The alloy becomes topologically nontrivial when the bands are inverted and $L_6^-$ ($L_6^+$) becomes the valence (conduction) band. [8] Figure 1(b) shows the band inversion versus Sn content, which occurs at $x_c \sim 0.16$ at 4.2 K. [3]

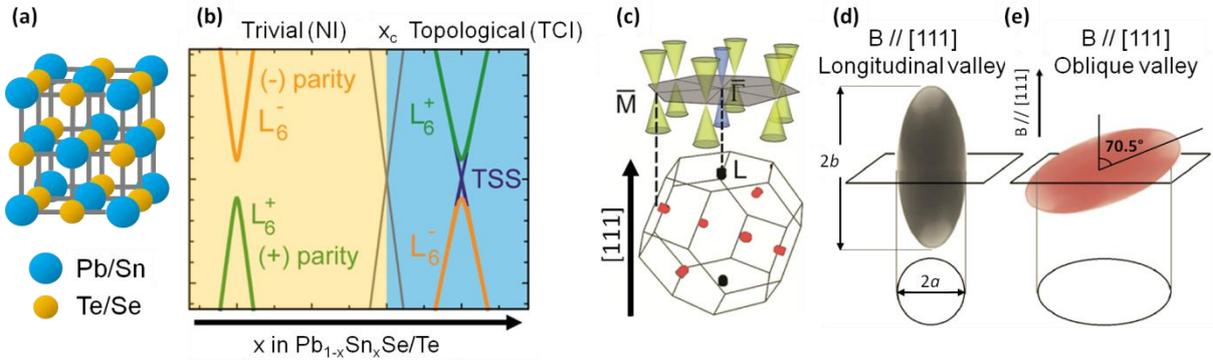

**Figure 1. (a)** Rocksalt crystalline structure of the lead salt compounds. In blue are the cations and in yellow the anions. **(b)** Composition-induced topological phase transition of $Pb_{1-x}Sn_xSe$ and $Pb_{1-x}Sn_xTe$ compounds from Normal Insulating (NI) regime (orange shaded) to TCI phase (blue). The band parity at the four $L$-points is inverted in the topological regime ($x \gtrsim 0.16$ at 4.2 K for $Pb_{1-x}Sn_xSe$). **(c)** The first Brillouin zone of the rocksalt structure with its 2D projection onto the studied (111) surface. In the topological regime, four Dirac cones occur at the four $L$-points. In (111) orientation, three $L$-points are oriented oblique to the surface (oblique valleys, red), and the other one is perpendicular to the surface (longitudinal valley, black). The constant-energy ellipsoids are represented in **(d)** for longitudinal valley and **(e)** for oblique valleys. The main ellipsoid axis of the longitudinal valley is parallel to [111] direction while for oblique valleys it is tilted by 70.5°.

As the band edge maxima occur at the $L$-point of the Brillouin zone, the Fermi surface anisotropy needs to be taken into account. The bulk Fermi surface shown in Fig. 1(c-e) is degenerate. It consists of one ellipsoid oriented parallel to the [111] direction, referred to as the longitudinal valley (Fig. 1(d)) and three tilted ellipsoids known as the oblique valleys (Fig. 1(e)). The ellipsoids are characterized by the anisotropy factor $K = (b/a)^2$, where $b$ and $a$ are the major and minor axis respectively, as illustrated in Fig. 1(d).

## I. <u>Experimental method</u>

We perform magneto-optical spectroscopy at different temperatures and magnetic fields on (111) $Pb_{1-x}Sn_xSe$ films of 2 μm thickness grown by molecular beam epitaxy on (111) $BaF_2$ substrates, as described in our previous work. [3] [24] The sample composition was precisely determined by X-ray diffraction (XRD) using the Vegard's law (Appendix A) [3] and the carrier concentration is in the low $10^{17}$ cm$^{-3}$ range. [18] At 4.2 K, far and mid-infrared (FIR and MIR) absorption spectra were recorded with the samples mounted on a probe coupled to a FTIR spectrometer. The absorption was measured using a composite Si-bolometer mounted just below the sample and cooled to 4.2 K. For temperature dependent experiments up to 200 K, we use either an external HgCdTe detector in the MIR range (80 to 300 meV) or an external Si-bolometer in the FIR range (from 10 to 80 meV). A superconducting coil provides a magnetic field up to 17 T parallel to the [111] direction in Faraday geometry.

## II. Model for Landau levels

In order to interpret the magneto-optical data, we used a 4-band $\boldsymbol{k}.\boldsymbol{p}$ model derived in Appendix B. The model is solved at $k_z = 0$, where the joint density of state is extremal. In the Faraday geometry, selection rules correspond to $\Delta N = \pm 1$ and the conservation of the effective spin. [1] [25] Thus, one obtains the energies of the magneto-optical interband transitions:

$$E_{N^c} - E_{N^v} = \sqrt{\Delta^2 + 2v_D^2 \hbar eBN^c + \left(\frac{\hbar eBN^c}{\widetilde{m}}\right)^2} + \sqrt{\Delta^2 + 2v_D^2 \hbar eBN^v + \left(\frac{\hbar eBN^v}{\widetilde{m}}\right)^2} \qquad (1)$$

where $2\Delta$ denotes the band gap energy. $N^{c,v}$=0, 1, 2, ... is the LL index of the conduction and valence band respectively. The far-bands interactions with $L_6^+$ and $L_6^-$ are represented by the effective mass $\widetilde{m}$. Remarkably, aside from the $B^2$-term, Eq. (1) accounts for transitions between Dirac Landau levels with the Dirac velocity $v_D$ given by [3]:

$$v_D^2 = v_c^2 + \frac{\Delta}{\widetilde{m}} \qquad (2)$$

This parameter includes exactly the interaction between $L_6^+$ and $L_6^-$ (term $v_c$) and incorporates the far-bands in a $k^2$ approximation (term $\Delta/\widetilde{m}$). In the following, $v_D^o$ ($v_D^l$) will refer to the oblique (longitudinal) valleys. We can then define the Dirac mass (i.e. the positive band-edge mass of $L_6^\pm$) as $m_D^{l,o} = \Delta/v_D^{l,o\,2}$. This is equivalent to defining the Dirac mass as: [26]

$$\frac{1}{m_D} = \frac{v_c^2}{\Delta} + \frac{1}{\widetilde{m}}$$

Thus, the LL energies are described by three parameters: $\Delta$, $v_D^{o,l}$, and $\widetilde{m}$. As the far-bands lie more than $1eV$ above and below $L_6^-$ and $L_6^+$, we consider them as independent of $T$ and $x$, meaning that $\widetilde{m}(x,T)$ is constant. [27] [28] For PbSe we use the values $\widetilde{m} = 0.23m_0$ for the longitudinal valley and $\widetilde{m} = 0.28m_0$ for the oblique ones as reported in Ref. [1] [29]. Finally, magneto-optical interband transitions yield the two Dirac parameters: $\Delta(x,T)$ and $v_D^{l,o}(x,T)$.

## III. Composition-induced topological phase transition

Representative transmission spectra in the MIR range at 4.2 K are shown in Fig. 2(a-d) at different magnetic fields for four samples having $0.10 < x < 0.25$. Absorptions due to the oblique or longitudinal valleys are marked by red and black arrows respectively. They mark a noticeable splitting of the absorption lines into two of unequal amplitude. The oblique valleys being threefold degenerate contribute a stronger absorption. This is consistently observed in Fig. 2(a-d) where the red arrows show systematically stronger absorption than black arrows. This valley splitting disappears for high Sn concentrations $x > 0.2$. The origin of this effect is discussed later in detail. The absorptions for both valleys are compared to Eq. (1) obtained from the $\bm{k}.\bm{p}$ model. The comparison leads to Fig. 2(e-h) where experimental magneto-optical absorption energies are plotted as symbols versus magnetic field and the fit to the $\bm{k}.\bm{p}$ model is represented by the solid lines. Agreement is excellent and allows us to accurately determine the energy gap $2|\Delta(x)|$ and the Dirac velocity $v_D^{L,o}(x)$ independently for both valleys.

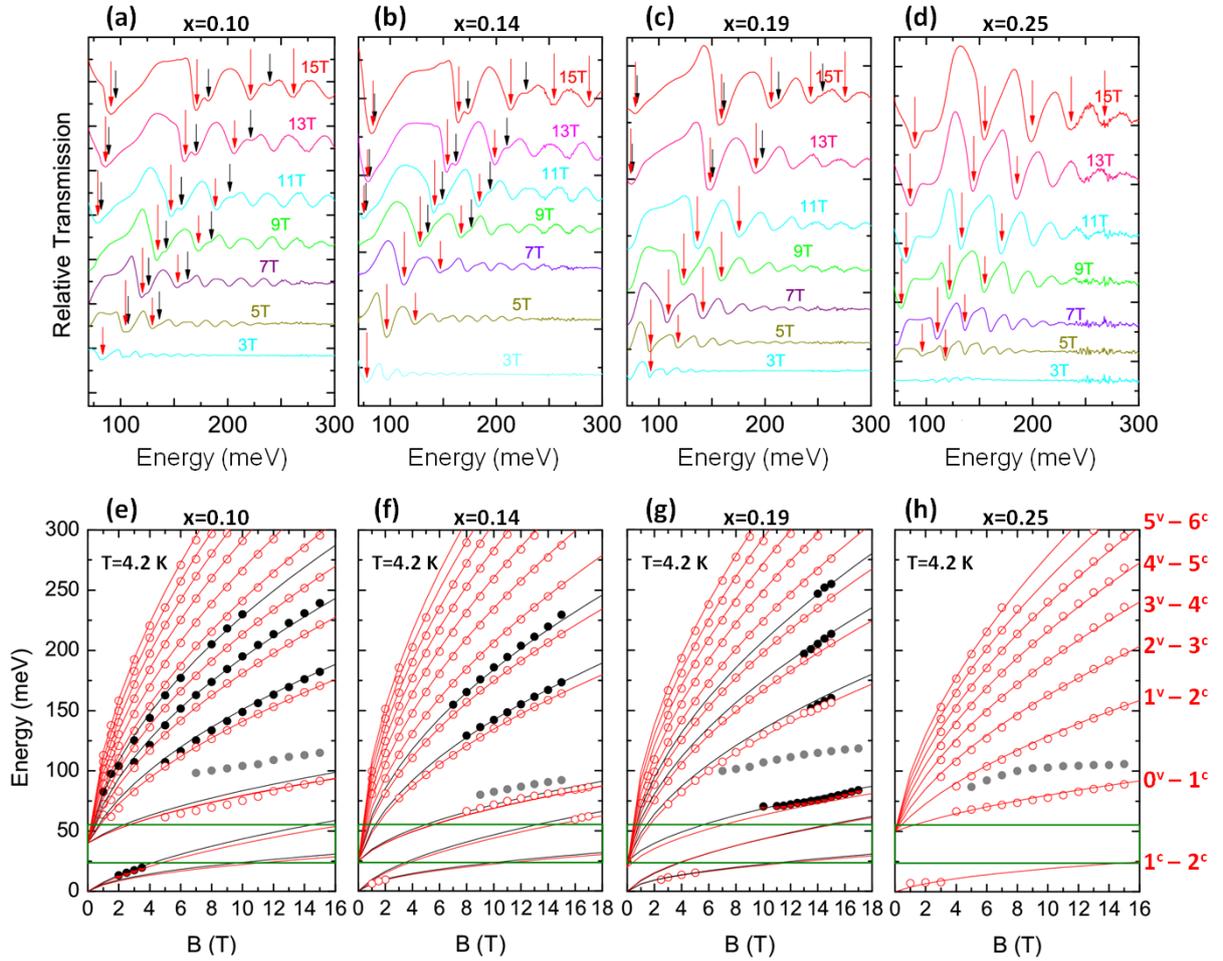

**Figure 2. (a-d)** Transmission spectra of Pb$_{1-x}$Sn$_x$Se with $x$ ranging from 0.10 to 0.25 at different magnetic fields normalized by the spectra at $B = 0$. Curves are shifted for clarity. Transmission minima, marked by black and red arrows, indicate optical absorption between Landau levels. They are represented by dots in the corresponding Landau fan charts depicted in **(e-h)**, where the experimental data (symbols) are fitted by solid lines corresponding to the massive Dirac model of Eq. (1). Transitions are labeled on the right by N$^{c,v}$-N+1$^{c,v}$ (N: LL index; c or v are for conduction or valence band). Red stands for oblique valleys and black for longitudinal valley. The gray dots are experimental absorptions attributed to the interband 1$^v$-0$^c$ transition shifted at $k_z \neq 0$ due to the populated 0$^c$ level. The green rectangles denote the restrahlen absorption band of the substrate BaF$_2$.

**Table I.** Experimentally derived Dirac parameters for the measured Pb$_{1-x}$Sn$_x$Se samples at $4.2\ K$. Uncertainties are $\pm 5$ meV for $2\Delta$, $\pm 0.05 \times 10^5$ m/s for $v_D$, $\pm 0.002 m_0$ for the Dirac masses, and $\pm 0.05$ for the anisotropy factor $K$.

| Name | $x$ | $2\Delta$ (meV) | $v_D^o$ (m/s) | $v_D^l$ (m/s) | $m_D^o/m_0$ | $m_D^l/m_0$ | $K$ |
|---|---|---|---|---|---|---|---|
| MBE-00 | 0.00 | +145 | 5.60 x 10$^5$ | 6.45 x 10$^5$ | 0.041 | 0.031 | 1.94 |
| MBE-05 | 0.05 | +85  | 5.20 x 10$^5$ | 5.80 x 10$^5$ | 0.028 | 0.022 | 1.66 |
| MBE-10 | 0.10 | +40  | 4.85 x 10$^5$ | 5.25 x 10$^5$ | 0.015 | 0.013 | 1.44 |
| MBE-14 | 0.14 | +25  | 4.75 x 10$^5$ | 4.90 x 10$^5$ | 0.012 | 0.011 | 1.15 |
| MBE-16 | 0.16 | −10  | 4.70 x 10$^5$ | 4.75 x 10$^5$ | 0.004 | 0.004 | 1.05 |
| MBE-19 | 0.19 | −20  | 4.55 x 10$^5$ | 4.60 x 10$^5$ | 0.008 | 0.008 | 1.05 |
| MBE-25 | 0.25 | −50  | 4.30 x 10$^5$ | 4.30 x 10$^5$ | 0.024 | 0.024 | 1.00 |
| MBE-28 | 0.28 | −80  | 4.10 x 10$^5$ | 4.10 x 10$^5$ | 0.042 | 0.042 | 1.00 |
| MBE-30 | 0.30 | −100 | 3.95 x 10$^5$ | 3.95 x 10$^5$ | 0.056 | 0.056 | 1.00 |

The derived energy gap $2\Delta$ and the Dirac velocities $v_D^{l,o}$ of the investigated samples are listed in Table I, and the corresponding Dirac masses and the anisotropy factor are shown as well. Figure 3(a) shows the variation of the energy gap $2|\Delta|$ versus composition at 4.2 K. $2|\Delta|$ decreases for $x < x_c$ and increases after reaching $x_c$. The minimal value of $2|\Delta|$ marks the change of topology around a critical composition $x_c \sim 0.16$. Thus, the sign of $2\Delta(x)$ becomes negative when entering the topological phase shaded in blue. The experimental dependence of $2\Delta(x)$ for $0 < x < 0.3$ is represented in Fig. 3(b). It can be well fitted by the following empirical formula that takes into account the sign change (dashed line in Fig. 3(b)):

$$\Delta(x) = 72 - 710x + 2440x^2 - 4750x^3 \quad \text{(meV)} \qquad (3)$$

Eq. (3) differs from previous experimental laws, which were linear in $x$. [9] [14] The high order polynomial expression accounts for the saturation of the energy gap near the composition-induced topological phase transition. Magneto-optical data are accurate enough to highlight this phenomenon. Such a non-linear variation was for instance predicted by recent tight-binding calculations in Pb$_{1-x}$Sn$_x$Te as being caused by alloy disorder. [30]

Note that Eq. (3) represents the ideal case under the assumption that the gap closes at $x_c$. However, as seen in Fig. 3(b) the gap does not completely close at the critical point, but a jump from positive to negative values is observed. This effect can be taken into account by modeling an avoided-crossing around $x_c$ between the bulk bands resulting in a renormalized gap $\widetilde{\Delta}(x)$: [29]

$$\widetilde{\Delta}(x) = \pm\sqrt{\Delta^2(x) + W^2} \qquad (4)$$

where $W = 7$ meV represents the non-closing gap at the critical point, and the $\pm$ sign stands for trivial and topological regime respectively. The solid lines in Fig. 3(b) represent the calculated $\widetilde{\Delta}(x)$ using Eq. (4). The origin of the non-closure of the energy gap is discussed in section IV.

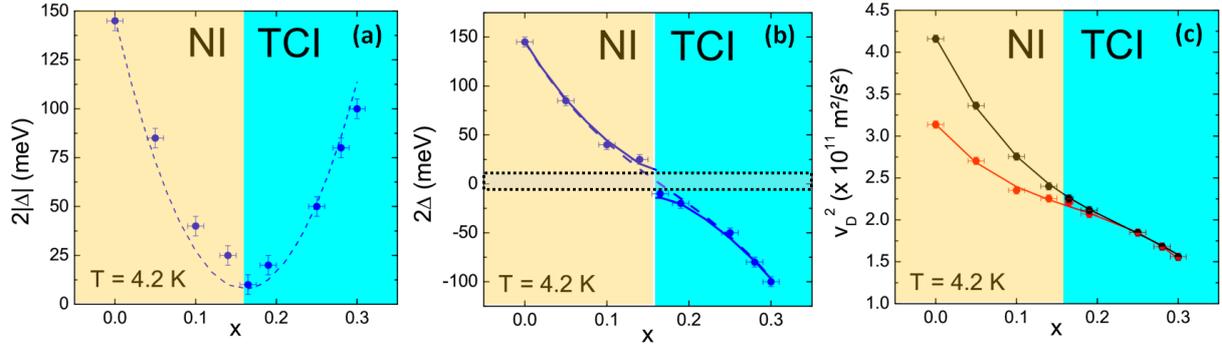

**Figure 3. (a)** Variation of the energy band gap modulus of $Pb_{1-x}Sn_xSe$ with $x$. The dashed line is a guide-for-the-eye, from which the minimum value indicates the transition to the TCI regime ($x = x_c$). In **(b)** the energy band gap is plotted versus Sn content. Experimental data are fitted with Eq. (3) (dashed line) and Eq. (4) (solid lines) which takes into account the jump of the energy gap across the topological phase transition. **(c)** Dirac velocity for the oblique (red) and the longitudinal valleys (black) versus Sn content. Solid lines are the fits from Eq. (5).

In Fig. 3(c), the experimental square Dirac velocities $v_D^l$ and $v_D^o$ are plotted and fitted versus Sn content. We can find the following empirical laws for the variation of the velocity:

$$\begin{cases} v_D^o(x) = \sqrt{0.314 - 1.13\,x + 4.79\,x^2 - 9.34\,x^3} & (10^6 \text{ m/s}) \\ v_D^l(x) = \sqrt{0.417 - 1.95\,x + 6.28\,x^2 - 8.84\,x^3} & (10^6 \text{ m/s}) \end{cases} \quad (5)$$

The ratio of $v_D^l(x)$ to $v_D^o(x)$ leads to the determination of the anisotropy factor $K(x)$ without the need to perform any angular dependent measurement (see Appendix C). $K(x)$ is reported in Table I. Remarkably, for $0.20 < x < 0.30$ the anisotropy becomes very weak (Fig. 3(c)) as the valleys become spherical and thus identical within experimental resolution. This accounts for the suppression of the valley splitting observed in the magnetooptical spectra in Fig. 2(d). Thus for $0.2 < x < 0.3$, $Pb_{1-x}Sn_xSe$ is a massive 3D Dirac material with mirror symmetric conduction and valence bands, an inverted band structure at crystalline symmetric points, and a nearly spherical Fermi surface. It is, in our view, a textbook model system to study topological phases of matter based on the Dirac equation.

# IV. Temperature-induced topological phase transition

Next, we have performed magneto-optical LL infrared spectroscopy at different temperatures between 4.2 K and 200 K. MIR spectra for the samples with $x = 0.19$ (MBE-19) and $x = 0.25$ (MBE-25) are shown in Fig. 4(a) and 4(b) respectively and three additional samples with $x = 0.10, 0.14$ and $0.28$ were investigated as well.

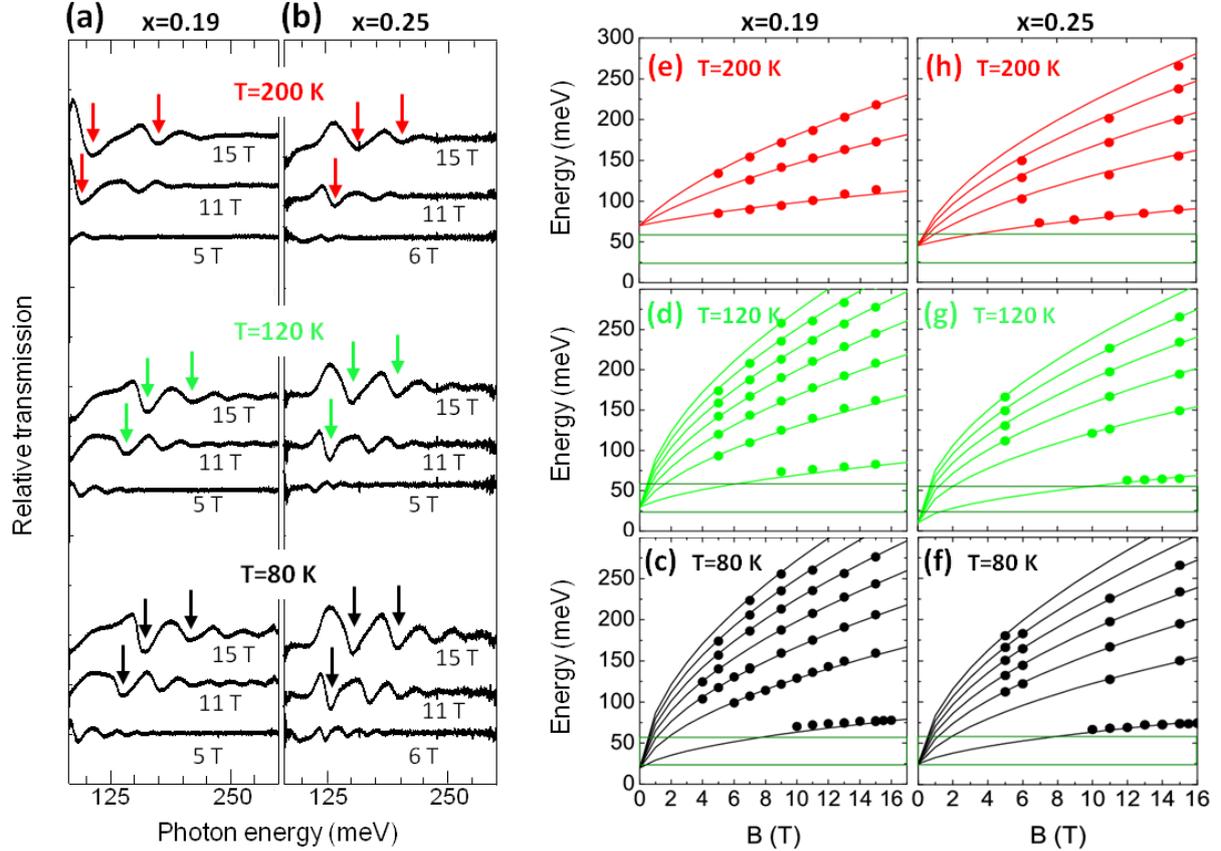

**Figure 4.** Transmission spectra in the MIR range normalized by the spectra at $B = 0$ of Pb$_{1-x}$Sn$_x$Se with $x = 0.19$ **(a)** and $x = 0.25$ **(b)** at three different magnetic fields and temperatures (80, 120, and 200 K). The corresponding fan charts are plotted in **(c-e)** and **(f-h)**, respectively. Experimental data (symbols) are fitted by solid lines corresponding to Eq. (1). Black, green, and red colors stand for the results obtained at 80, 120, and 200 K respectively.

Sample mobilities are sufficiently high to resolve the LLs up to 200 K. Thus, using the $\mathbf{k}\cdot\mathbf{p}$ description of the LLs, $\Delta(T)$ and $v_D^o(T)$ can be extracted at each temperature from the fits presented in Fig. 4(c-h). The slight broadening of the Landau levels makes the valley anisotropy difficult to resolve at elevated temperature above 80 K as clearly seen in Fig. 4(a). The analysis is thus restricted to the oblique valley. An accurate determination of $\Delta$ at each temperature is possible and leads to the following phenomenological formula:

$$\Delta(x,T) = \Delta(x, 4.2) - 22.5 + \sqrt{22.5^2 + 0.32^2(T - 4.2)^2} \quad \text{(meV)} \quad (6)$$

where $\Delta(x, 4.2)$ is given by Eq. (3). This dependence, valid for $0 < x < 0.3$ and $4.2\ K < T < 200\ K$, is more accurate than the one previously derived by Strauss *et al.* [9]. Experimental data of $\Delta(x,T)$ (symbols) as well as the fit using Eq. (6) (solid lines) are summarized in Fig. 5(a).

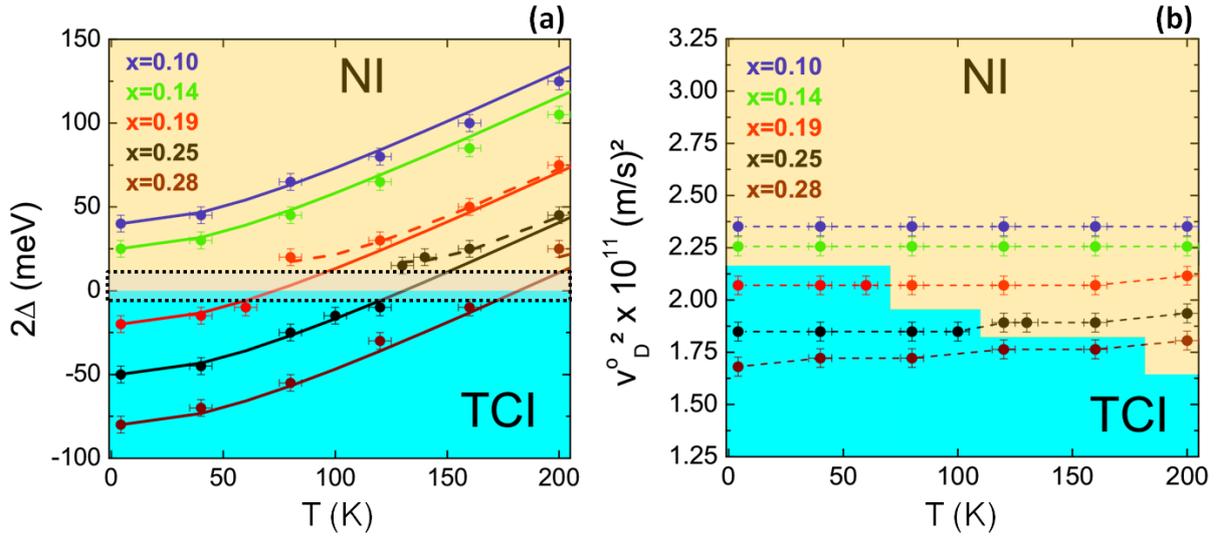

**Figure 5. (a)** Energy gap variation with temperature of $Pb_{1-x}Sn_xSe$ for various Sn content $x$ represented in different colors. Data (symbols) are fitted with Eq. (6) (solid lines). Guide-for-the-eyes are shown in dashed lines to highlight the jump of the energy gap across the temperature-induced topological phase transition, represented by the blue and orange regions. **(b)** Square Dirac velocities plotted versus temperature for different Sn content $x$ represented in the same colors than in **(a)**. Dashed lines are guide-for-the-eyes.

Note that $\Delta(x,T)$ is almost linear for $T > 40$ K and below this temperature, it is nearly constant with $T$. Such a gap saturation has been previously reported by Wojek *et al.* using ARPES, which however is more sensitive to the electronic structure of the surface and not that of the bulk. [14] We can clearly see that $\Delta$ increases with temperature. Thus, if the energy gap is negative at 4.2 K at $x > x_c$, it will turn positive at a higher temperature depending on the composition.

Remarkably, for the topological samples ($x > x_c$), a non-closure of the energy gap in the vicinity of the phase transition is observed (gray rectangle in Fig. 5(a)). As a matter of fact, for all three samples with $x = 0.19, 0.25$ and $0.28$ the data points within this transition region consistently deviate from the fitted solid lines from Eq. (6), i.e., a jump from negative to positive energy gap is observed near the critical temperature of the phase transition for all compositions. Guide-for-the-eyes illustrate this effect in Fig. 5(a). This phenomenon is similar to the one observed for the composition-induced topological phase transition in Fig. 3(b).

Since in our experiments we probe the bulk band structure, this observation confirms that the absence of gap closure previously reported in Ref. [14] is of bulk origin. Recent density function theory calculations [30] proposed that this effect may originate from alloy disorder. Indeed, such disorder could lead to an avoided-crossing between conduction and valence bands near the critical point in temperature and magnetic field. [29] Although it remains open what mechanism causes this energy-gap jump, our experimental verification as a function of composition and temperature independently for several samples highlights that it is a very well reproducible effect, which can affect the nature of the topological phase transition.

The measured square Dirac velocity $v_D^{o\,2}(x,T)$ is represented in Fig. 5(b). Data obtained between 4.2 K and 200 K are plotted for each composition. For a given $x$, $v_D^{o\,2}(x,T)$ is nearly constant with temperature, similar as what has been reported in $Hg_{1-x}Cd_xTe$ crystals close to the critical composition. [31] We have thus experimentally obtained a complete determination of the Dirac parameters as a function of $x$ and $T$ as summarized in Fig. 5 for $\Delta(x,T)$ and $v_D(x,T)$.

## V. Topological phase diagram

The quantitative determination of the Dirac parameters of $Pb_{1-x}Sn_xSe$ versus temperature and composition for $0.10 \leq x \leq 0.28$ allows us to construct the temperature-composition topological phase diagram presented in Fig. 6. If the transition is assumed to be sharp with a well-defined 3D zero-gap state at the phase boundary, this boundary can be simply computed by setting Eq. (6) equal to zero. This yields the red phase boundary shown in Fig. 6. Nevertheless, one has not been able to observe a stable 3D zero-gap state at the phase boundary. [16] [29] The question of whether or not this state is stable in 3D has been theoretically treated under different perspectives. [30] [32] With the phase diagram established we motivate future work to elucidate the origin of the energy-gap jump around the topological phase transition.

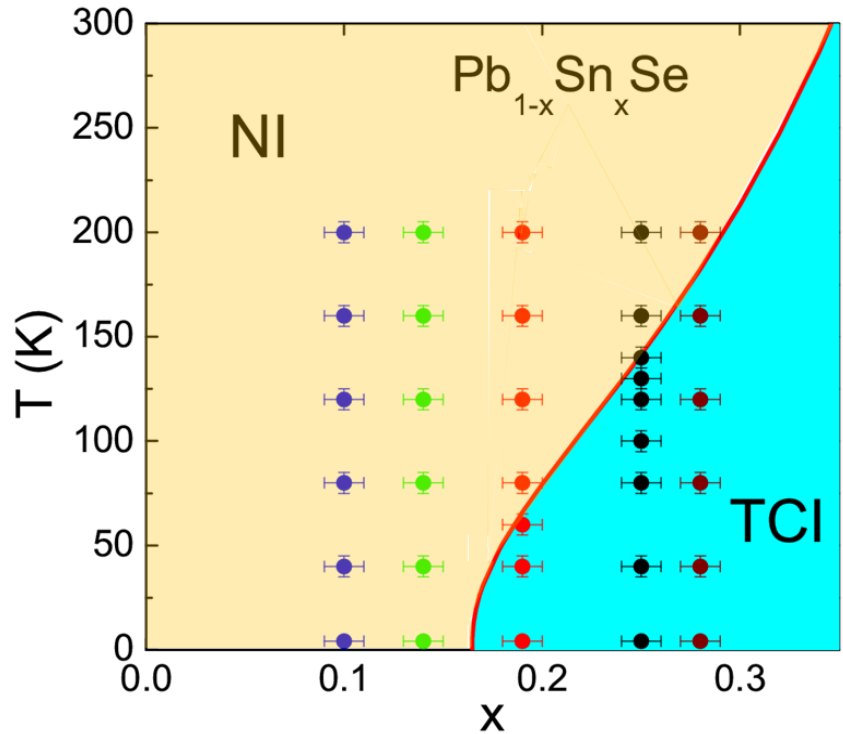

**Figure 6.** Topological phase diagram of $Pb_{1-x}Sn_xSe$ versus temperature and Sn content. Dots represent the measured samples with different compositions (in colors). The topological regime is represented in blue and delimited from the trivial regime in orange by the red line calculated from Eq. (6).

## V. Conclusion

In conclusion, we have systematically derived the Dirac bulk band parameters of $Pb_{1-x}Sn_xSe$ for a wide range of compositions and temperatures using magneto-optical experiments, providing accurate values of the energy gap and Dirac velocities in dependence of $x$ and $T$. These band parameters are crucial for an accurate description of the electronic properties of the topological system. From our results we establish the topological phase diagram as a function of temperature and composition. Most notable, we consistently observe that the fundamental gap does not close completely to zero across the topological phase transition. Our result motivates further theoretical considerations to understand the evolution of the bulk band structure around the critical composition and temperature. This is not only of fundamental interest but also of key importance for device applications based on tuning the topological properties of the system by external means like strain or applied electric fields. [21] [22] [23]

**AKNOWLEDGMENTS**

We acknowledge fruitful discussions with G. Bastard and R. Ferreira. This work is supported by Agence Nationale de la Recherche LabEx ENS-ICFP Grant No ANR-10-LABX-0010/ANR-10-IDEX-0001-02 PSL and by the Austrian Science Fund, Projects P 28185-N27 and P 29630-N27. G.K. is also partly supported by a PSL scholarship.

## APPENDIX A: SAMPLE GROWTH AND CHARACTERIZATION

In order to study the topological transition, fully relaxed 2 μm thick $Pb_{1-x}Sn_xSe$ epilayers were grown by MBE on $BaF_2$ (111) substrates. [33] [34] The films were characterized by *in-situ* reflection high energy electron diffraction and *ex-situ* atomic force microscopy. The sample characterization evidences the complete relaxation of the lattice mismatch of the epilayers with respect to its substrate $BaF_2$. [3] XRD is used in order to determine the lattice parameter with a precision of $\pm 10^{-3}$ Å (see Ref. [3]). Figure 7 shows the evolution with $x$ of the lattice parameter determined in XRD. It displays a linear variation described by the following Vegard's law:

$$a(x) = -0.1246x + 6.1240 \quad (\text{Å})$$

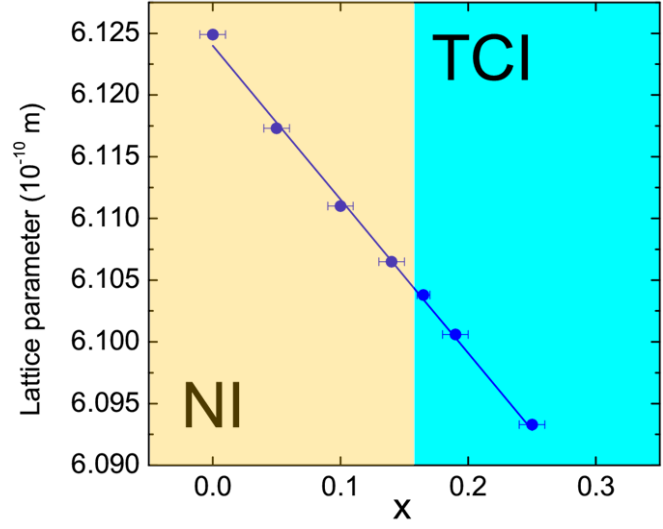

**Figure 7.** Vegard's law for $Pb_{1-x}Sn_xSe$ epilayers grown on $BaF_2$ (111) substrates. The experimental value of the lattice parameter between $x = 0$ and $x = 0.3$ gives the Sn to Pb ratio.

Transport measurements were performed to extract the carrier density and mobility. Carrier densities as low as $10^{17}$ cm$^{-3}$ are achieved. The Hall mobility μ of the samples at 77 K is measured to be around 30000 to 60000 cm²/Vs. [18] These excellent transport properties allow us to observe Landau quantization at low magnetic fields and high temperature.

## APPENDIX B: $k.p$ MODEL

A 4 band $k.p$ theory has been developed by Mitchell & Wallis to obtain the $L_6^{\pm}$ LLs of the lead salts. [35] The $Pb_{1-x}Sn_xSe$ bulk system in a (111)-oriented magnetic field ($\vec{B}//z$) is described by the following Hamiltonian [25] [28]:

$$\begin{pmatrix} \Delta - \left(N - \frac{1}{2}\right)\hbar\widetilde{\omega}_c - \frac{1}{2}g_c\mu_B B & 0 & \hbar v_z k_z & \sqrt{2e\hbar v_c^2 BN} \\ 0 & \Delta - \left(N + \frac{1}{2}\right)\hbar\widetilde{\omega}_c + \frac{1}{2}g_c\mu_B B & \sqrt{2e\hbar v_c^2 BN} & -\hbar v_z k_z \\ \hbar v_z k_z & \sqrt{2e\hbar v_c^2 BN} & -\Delta + \left(N - \frac{1}{2}\right)\hbar\widetilde{\omega}_v + \frac{1}{2}g_v\mu_B B & 0 \\ \sqrt{2e\hbar v_c^2 BN} & -\hbar v_z k_z & 0 & -\Delta + \left(N + \frac{1}{2}\right)\hbar\widetilde{\omega}_v - \frac{1}{2}g_v\mu_B B \end{pmatrix}$$

Here $v_c$ and $v_z$ are given by the interband momentum matrix element for direction perpendicular and parallel to [111]. $\widetilde{\omega}_{c,v}$ and $g_{c,v}$ represent the far-band contributions to the effective mass and g-factor of $L_6^\pm$. [1] [18]

Since the far-bands are nearly equally distant from $L_6^\pm$, we make the following assumptions [18]:

$$\begin{cases} \widetilde{\omega}_c = -\widetilde{\omega}_v = \widetilde{\omega} = \dfrac{eB}{\widetilde{m}} \\ g_c = -g_v = g \\ g\mu_B B = -\hbar\widetilde{\omega} \end{cases}$$

and the Hamiltonian simply writes:

$$\begin{pmatrix} \Delta - (N-1)\hbar\widetilde{\omega} & 0 & \hbar v_z k_z & \sqrt{2e\hbar v_c^2 BN} \\ 0 & \Delta - (N+1)\hbar\widetilde{\omega} & \sqrt{2e\hbar v_c^2 BN} & -\hbar v_z k_z \\ \hbar v_z k_z & \sqrt{2e\hbar v_c^2 BN} & -\Delta + (N-1)\hbar\widetilde{\omega} & 0 \\ \sqrt{2e\hbar v_c^2 BN} & -\hbar v_z k_z & 0 & -\Delta + (N+1)\hbar\widetilde{\omega} \end{pmatrix}$$

This Hamiltonian can be solved exactly at $k_z = 0$ and gives the LL energies for B//[111]. [1] [25] [35]

## APPENDIX C: ANISOTROPY

For a (111)-oriented sample (Fig 1(c)), the longitudinal valley consists of a bulk carrier ellipsoidal pocket having its major axis parallel to the [111]-direction (Fig. 1(d)). The three oblique valleys have their major axis tilted by $\theta = 70.5°$ with respect to the [111]-direction (Fig. 1(e)).

We assume that the bulk carriers in the two valleys are described by the same Hamiltonian developed just above. For a magnetic field B//[111], the expressions of the LL energies remain unchanged for oblique valleys as far as we consider that $v_D$ depends on $\theta$. Using the expression of the cyclotron frequency in a tilted valley given by Pascher *et al* [26], one can derive the Dirac velocity for a given angle $\theta$:

$$v_D^{tilted}(\theta) = v_D^l \left(\frac{1}{K}\sin^2\theta + \cos^2\theta\right)^{1/4}$$

where $K$ is the anisotropy factor of the Fermi surface. $K$ is defined as the ratio between the great and minor axis of the ellipsoid $K = (b/a)^2$ (see Fig. 1(d-e)) or equivalently as $(v_c/v_z)^2$.

As $\theta = 70.5°$ for the oblique valleys, one gets:

$$v_D^o = v_D^l \left(\frac{8}{9K} + \frac{1}{9}\right)^{1/4} \qquad (7)$$

The evolution of the anisotropy $K(x)$ in Pb$_{1-x}$Sn$_x$Se is thus deduced from the experimental determination of $v_D^{l,o}(x)$ (see Table I).

Finally, the determination of $K(x)$ and $v_c(x)$ from Eq. (7) and Eq. (2) leads to $v_z(x)$. Thus, all parameters of the Hamiltonian in Appendix B are experimentally determined.